# Field-free spin-orbit torque switching from geometrical domain wall pinning

*Jong Min Lee,† Kaiming Cai,† Guang Yang, Yang Liu, Rajagopalan Ramaswamy, Pan He, and Hyunsoo Yang\**

Department of Electrical and Computer Engineering, National University of Singapore, 117576, Singapore

**ABSTRACT:** Spin-orbit torques, which utilize spin currents arising from the spin-orbit coupling, offer a novel method to electrically switch the magnetization with perpendicular anisotropy. However, the necessity of an external magnetic field to achieve a deterministic switching is an obstacle for realizing practical spin-orbit torque devices with all-electric operation. Here, we report a field-free spin-orbit torque switching by exploiting the domain wall motion in an anti-notched microwire with perpendicular anisotropy, which exhibits multi-domain states stabilized by the domain wall surface tension. The combination of spin-orbit torque, Dzyaloshinskii-Moriya interaction, and domain wall surface tension induced geometrical pinning allows a deterministic control of the domain wall and offers a novel method to achieve a field-free spin-orbit torque switching. Our work demonstrates the proof of concept of a perpendicular memory cell which can be readily adopted in a three-terminal magnetic memory.

The spin-orbit torque (SOT) driven switching of the magnetization with perpendicular magnetic anisotropy (PMA) is receiving immense research attention[1-3] as it can be realized into a three-terminal magnetic memory.[4, 5] The key advantage of a three-terminal magnetic memory is the separation of read and write current paths as the conventional two-terminal spin-transfer torque (STT) memory suffers from the read disturbance and tunnel barrier breakdown.[6] However, an additional external magnetic field is required to achieve a deterministic SOT switching, which is an obstacle for practical applications. To tackle this issue, a field-free deterministic SOT switching was reported in literatures by utilizing a lateral wedge oxide,[7] tilted magnetic easy axis,[8] antiferromagnetic layer,[9, 10] or hybrid ferromagnetic/ferroelectric structure.[11]

On the other hand, an alternative three-terminal magnetic memory was reported using the STT-driven domain wall (DW) motion,[12, 13] where the switching is achieved by moving a DW back and forth along the magnetic wire. The key requirement of the DW motion device is to initialize the magnetization at the two ends of wire in opposite directions. As a result of the antiparallel state at the two ends, a single DW always exists in the wire which is essential to achieve successive switching processes by back and forth DW motion. Although STT-DW motion switching scheme offers a novel method to switch the magnetization, it involves complex material engineering and lithography processes to achieve the antiparallel state at the two ends, such as deposition of two additional magnetic layers with dissimilar anisotropies on the same height.[12, 13]

In this work, we demonstrate a field-free SOT switching by exploiting the SOT-DW motion in a PMA anti-notched microwire. In contrast to the complicated structural requirement of the STT-DW motion switching scheme, we use a simple Ta/CoFeB/MgO structure to realize the SOT-DW motion switching, which can be readily adopted in a conventional magnetic tunnel junction. The reliable and deterministic control of a single DW injection and DW displacement are obtained simply by sweeping the current, which enables a field-free SOT switching under repeated bipolar currents. The results are explained by the combination of the SOT-DW motion,[14-17] the Dzyaloshinskii-Moriya interaction,[18, 19] and the geometric domain wall pinning due to surface tension.[20] The proposed scheme can be easily integrated into three-terminal memory devices.

For this study, a Ta (6 nm)/Co$_{40}$Fe$_{40}$B$_{20}$ (1.1 nm)/MgO (2 nm)/SiO$_2$ (3 nm) structure was grown by sputtering and annealed at 250 °C for 30 minutes to improve PMA. As shown in Figure 1a, the film was patterned into an anti-notched microwire and the electrodes of Ta (5 nm)/Cu (100 nm) were deposited at the ends for electrical contacts. The width of microwire is 5 μm and the diameter of discs is 10 μm. The location between the straight microwire and the disc serves as a geometric pinning site as discussed later. A current ($I$) is applied along the microwire (along the $x$-axis) and an external magnetic field ($H$) is applied in the $xz$-plane with a polar angle of $\theta_H$ estimated by the relation, $H_{sw}(0) = H_{sw}(\theta_H)\cos\theta_H$,[21] where $H_{sw}(\theta_H)$ is the switching field at $\theta_H$.

A polar magneto-optical Kerr-effect (MOKE) microscope is used to capture the out-of-plane component of the magnetization ($M_z$) during the switching process. Each captured image is subtracted from a reference image taken at a magnetically saturated state along an opposite out-of-plane direction to obtain a differentiated MOKE image with an enhanced contrast. The contrast intensity from the differentiated MOKE image is averaged and normalized to quantify the average $M_z$ in the microwire. The quantified MOKE signal of +1 and –1 represents average $M_z$ along the +$z$-direction ($M_z > 0$) and –$z$-direction ($M_z < 0$), respectively. Figure 1b shows the measured MOKE signal from the patterned microwire under a perpendicular field. The square-shaped

MOKE signal indicates a clear PMA in the patterned structure. The insets of Figure 1b show the differentiated MOKE images with $M_z > 0$ (light contrast) and $M_z < 0$ (dark contrast).

The SOT switching in an anti-notched microwire is first studied in the presence of $H$. Figure 2a,b show the results when the in-plane $H$ is applied along the $+x$ ($\theta_H \sim 90°$) and $-x$-direction ($\theta_H \sim 270°$), respectively. The magnitude of in-plane magnetic field $H$ is fixed at 495 Oe and the current sweep rate is 0.1 mA/s. The switching sequence is reversed when $H$ is reversed and each switching curve is approximately symmetric with respect to $I$. Further, the observed switching directions are in agreement with previous SOT switching reports using a Ta channel.[2, 22, 23] The A and B in the inset of Figure 2a,b show the MOKE images before and after the switching events, respectively, for the case when $I$ is swept from negative to positive.

We then apply $H$ slightly tilted away from in-plane and the results shown in Figure 2c ($\theta_H \sim 88°$) and 2d ($\theta_H \sim 268°$). The tilted $H$ gives rise to an out-of-plane component of magnetic field ($H_z$). The switching directions with respect to $I$ and $H$ remain the same as in the case of the in-plane $H$ in Figure 2a,b. However, the switching curves under tilted $H$ are not symmetric with respect to $I$ and are shifted to the left. Moreover, intermediate multi-domain states (A-D in Figure 2c,d) are observed during the switching process in contrast to the sharp switching under the in-plane $H$ (A-B in Figure 2a,b). These above results of the SOT switching described in Figure 2 shed light on the various contributions to the SOT switching process in the anti-notched structure as discussed in the following paragraphs.

Recent micromagnetics studies,[24, 25] as well as spatial[22] and time-resolved [26, 27] experiments have revealed that the SOT switching in PMA structures is realized by a reverse domain nucleation and expansion rather than a coherent magnetization rotation.[28, 29] The SOT utilizes a spin-polarized current which can be considered as an equivalent field of $H_{SOT} = -\dfrac{\theta_{SH} \hbar J}{2|e|M_s t}(\hat{m} \times \hat{y})$, when the substantial magnetization is orthogonal to spin current direction.[21] $\theta_{SH}$ is the spin-Hall angle, $J$ is the applied current density, $\hbar$ is the reduced Plank's constant, $e$ is the electron charge, $M_s$ is the saturation magnetization, and $t$ is the ferromagnetic layer thickness. Here, $\hat{m}$ and $\hat{y}$ indicate the direction of the internal DW magnetization and spin current, respectively. The spin-Hall angle is measured as –0.09 in our sample using the harmonic technique.[30] It is noted that the damping-like component of SOT (related to the spin-Hall angle) is used to interpret the results and the role of the field-like component is neglected as the damping-like component is reported to govern the deterministic SOT-DW motion and the SOT switching processes in the static regime.[15, 21, 30]

Under collinear $I$ and $H$, the DW experiences an out-of-plane $H_{SOT}$, as the direction of internal DW magnetization is dominantly determined by the applied $H$ direction.[21] For example, applying a positive $I$ exerts a perpendicular $H_{SOT}$ in the +z-direction (–z-direction) on the DW when $H$ is applied along the +x-direction (–x-direction). The $H_{SOT}$ reverses its direction when $I$ is applied in opposite direction. Consequently, once a reverse domain is nucleated, the $H_{SOT}$ drives the DW and leads to the switching via domain expansion with an anti-clockwise switching loop when $H$ is applied along the +x-direction (Figure 2a) and with a clockwise switching loop under $H$ along the –x-direction (Figure 2b). For the case of tilted $H$, there is a shift in the switching loop. The tilting of $H$ does not change the switching sequence as the directions of the internal DW magnetization and $H_{SOT}$ are dominantly determined by the x-component of $H$ ($H_x$). The tilted $H$ shifts the switching loop as $H_z$ assists or hinders the $H_{SOT}$ driven switching process depending on the relative directions of $H_z$ and $H_{SOT}$.[22] When $\theta_H \sim 88°$ and $I > 0$, both the $H_{SOT}$ and $H_z$ are in the +z-direction resulting in a reduction of the switching current. However, the switching current increases when $\theta_H \sim 88°$ and $I < 0$, as the $H_{SOT}$ and $H_z$ are in the opposite directions. As a result, the entire switching loop in Figure 2c shifts to the left. For the case $\theta_H \sim 268°$, the directions of both $H_{SOT}$ and $H_z$ are reversed. Therefore, the switching loop is still shifted to the left in Figure 2d.

As discussed earlier, the SOT switching in PMA structures is initiated by reverse domain nucleation and followed by reverse domain expansion.[22, 26, 31] However, in the case of in-plane $H$, we observed a sharp transition of the magnetization without capturing the reverse domain nucleation nor the reverse domain expansion (Figure 2a,b) due to the long time scale of the measurement.[28] On the other hand, the reverse domain expansion across the microwire is clearly observed under the tilted $H$ as we sweep $I$ from negative to positive (Figure 2c,d). The intermediate multi-domain states are shown to be stabilized below the corresponding threshold currents to achieve complete switching. The stabilized multi-domain states indicate the presence of additional contributions to the SOT switching process in the anti-notched microwire other than $H_{SOT}$ as discussed below.

The details of the switching process with multi-domain states under the tilted $H$ are illustrated in Figure 3. Under a positive current $+I_1$, the switching process is initiated by the reverse domain nucleation at the left end of the microwire as sketched in Figure 3a. The nucleated reverse domain expands along the microwire under the SOT, but the domain wall is pinned as it reaches the neck part between the straight microwire and the disc as shown in Figure 3b (corresponds to 'B' in insets of Figure 2c,d). A relatively larger current density ($+I_3$) is needed to overcome the pinning of the anti-notch structure. It is observed that the reverse domain is pinned at the left part of the microwire in Figure 2c,d which suggest that the reverse domain nucleation position is fixed at the left end of the microwire.

First, we discuss the physics behind this fixed nucleation position at the left end of the microwire. In principle, the switching direction should be the same regardless of the nucleation position. For example, the SOT would result in a similar anti-clockwise (clockwise) switching loop when *H* is applied along the +*x*-direction (−*x*-direction) even if the reverse domain is nucleated at the right end of the microwire because the direction of $H_{SOT}$ is dominantly determined by *H* direction.[21] The fixed nucleation position indicates a stabilized chiral DW in a presence of Dzyaloshinskii-Moriya interaction (DMI).[22, 32]

The DMI is an antisymmetric exchange interaction which favors the noncollinear alignment of neighboring spins[18, 19] and is responsible for magnetic configurations,[33, 34] in particular the DW chirality.[15-17] The DMI energy is expressed as $E_{DMI} = -D_{ij} \cdot (S_i \times S_j)$, where $D_{ij}$ is the DMI vector, and $S_i$ and $S_j$ indicate spin vectors in neighboring sites *i* and *j*. The DMI with a positive (negative) sign energetically favors a DW with right-handed (left-handed) chirality. Since a local tilt of the DW magnetization is determined by $H_x$, a preferential nucleation at one edge is expected depending on the sign of $H_x$. The DMI sign of our Ta/CoFeB/MgO structure is identified to be positive because the reverse domain is always nucleated at the left end of the microwire. For example, the nucleated DW in Figure 2c (Figure 2d) is expected to have ↑→↓ (↓←↑) configuration due to $H_x > 0$ ($H_x < 0$) which follows the right-handed chirality. If the reverse domain is nucleated at the right end of the microwire, the DW in Figure 2c (Figure 2d) would follow the left-handed chirality with ↓→↑ (↑←↓) configuration at $H_x > 0$ ($H_x < 0$). Since we only observe the DWs with the right-handed chirality, the DMI in our Ta/CoFeB/MgO has a positive sign and is consistent with the previous studies with similar structures.[22, 35, 36]

Next, we discuss the pinning of reverse domain at the neck part of the anti-notched microstrip. Such a positional deterministic DW pinning is quite surprising as a DW pinning in an anti-notched wire was only reported in structures with in-plane anisotropy.[37, 38] In the case of the in-plane anti-notched wire, the different shape anisotropies in a straight wire and a disc part trigger a DW injection and pinning under applied $H$. However, the effect of shape anisotropy is negligible in our structure due to a strong PMA. Recently, the DW surface energy was reported to give rise to a geometrically induced pinning in PMA structures.[20] The DW surface energy, which can be understood as a surface energy from an elastic membrane, exerts a Laplace pressure ($P_{Laplace}$) on the DW and leads to spontaneous contraction of the DW radius. In the presence of an opposing pressure which tends to expand the DW radius, the DW can be stabilized in a semicircular form as sketched in Figure 3b.[20] In the studied structure, the SOT can be seen as the opposing pressure ($P_{SOT}$) as the SOT tries to expand the DW into the disc part. Therefore, the DW is pinned at the neck part and stabilized in a semicircular form when the $P_{SOT}$ is canceled out by the $P_{Laplace}$. The DW surface energy linearly scales with an inverse of the width of the wire. As the radius of curvature of the DW becomes smaller, the Laplace pressure on the DW can be higher, which provides a higher stability and stronger immunity to the external magnetic field.[20]

The DW is depinned as we increase $I$ since the $P_{SOT}$ overcomes the $P_{Laplace}$ and then the reverse domain expands along the microwire as sketched in Figure 3c. However, the DW is pinned again at the next neck part due to $P_{Laplace}$ as sketched in Figure 3d (corresponds to 'C' in insets of Figure 2c,d). On applying even larger $I$, the DW is depinned and then the full switching is achieved (corresponds to 'D' in insets of Figure 2c,d). Note that the pinned DW at the neck part loses its semicircular form and becomes flat upon removing $I$ as sketched in Figure 3e since $P_{Laplace}$ leads to a spontaneous contraction of the DW radius. As a result, the multi-domains states can be deterministically controlled by sweeping $I$ above the corresponding threshold values for the reverse domain nucleation and DW depinning. The intermediates states are only observed under the tilted $H$ and $I > 0$, as the $H_z$ lowers the nucleation current below the depinning current. The different depinning currents at the left and the right discs of the microwire are attributed to different demagnetization fields as the portion of the up to down domain in the microwire is changed.[20]

Using an initial antiparallel configuration in two disks, we then demonstrate a deterministic field-free SOT switching under repeated bipolar currents. The microwire requires an initialization using both $I$ and $H$ as illustrated in Figure 3, however, the initialization is required only once after fabrication and no external magnetic field is required during the operation. The initial state is achieved from a multi-domain state (corresponds to 'C' in inset of Figure 2c). Subsequently, a fixed magnitude of $I$ is applied without $H$ to achieve the field-free switching by moving the DW back and forth along the magnetic wire. As shown in Figure 4a, the deterministic SOT switching consistently occurs by applying series of the bipolar currents. The DW moves with applying currents, but is pinned and stabilized at the neck part as shown in Figure 4b. Note that the DW is flat without applying $I$ (#0 in Figure 4b) while it is in a semicircular form under $I$ (#1~4 in Figure 4b), which clearly indicates that the DW pinning under the current is induced by the surface energy. The DW moves along the current flow direction in the absence of $H$, which indicates the right-handed chirality of DW with a positive DMI[15, 31] and is consistent with DMI sign from the results in Figure 2.

Figure 5a shows a comparison of switching currents between the conventional SOT switching under $H$ and our field-free SOT switching using geometrical DW pinning. For the case of the conventional SOT switching, the switching current increases as we decrease the magnitude of in-plane $H$ and reaches to a extrapolated value of ~ 5 mA at $H = 0$. On the other hand, it is observed that the field-free SOT switching from geometrical pinning can switch back and forth between the two states at a much smaller switching current of 3 mA. Therefore, in addition to the field-free operation, our proposed SOT switching scheme operates at a lower power compared to the conventional SOT scheme, and helps in building power efficient three terminal magnetic memories. Figure 5b depicts a proposed three terminal magnetic memory using our field-free SOT switching scheme. The proposed three terminal magnetic memory is designed to have a magnetic wire which serves as a free layer and a magnetic tunnel junction which is composed of a tunnel barrier and a reference layer with a fixed magnetization direction. The write current applied along the microwire switches the central part of the wire using the SOT-DW motion. The magnetization at the two ends are not switched due to the surface energy from the antinotches. A small read current passes through the magnetic tunnel junction and senses a tunneling magnetoresistance which differs for the parallel ($R_\mathrm{P}$) and the antiparallel ($R_\mathrm{AP}$) alignments between the free layer and the reference layer. The initialization of the wire requires both $I$ and $H$ as illustrated in Figure 3, however, no external magnetic field is required during the operation. Note that multiple wires in an integrated chipset can be simultaneously initialized with standard commercial probe cards by applying a uniform magnetic field across the chipset area.

We have investigated the SOT switching behaviors in the Ta/CoFeB/MgO anti-notched microwire with PMA. The SOT switching was initiated by reverse domain nucleation at one edge followed by domain expansion across the microwire. Multi-domain states were stabilized during the switching process and showed a deterministic switching under SOTs. The multi-domain states were explained in terms of the DMI and the competition between the SOT equivalent field and the DW surface energy. Our work paves a novel way for a reliable and deterministic DW injection and manipulation in PMA structures. In addition, the demonstrated field-free SOT switching scheme can be readily incorporated into a three terminal magnetic memory

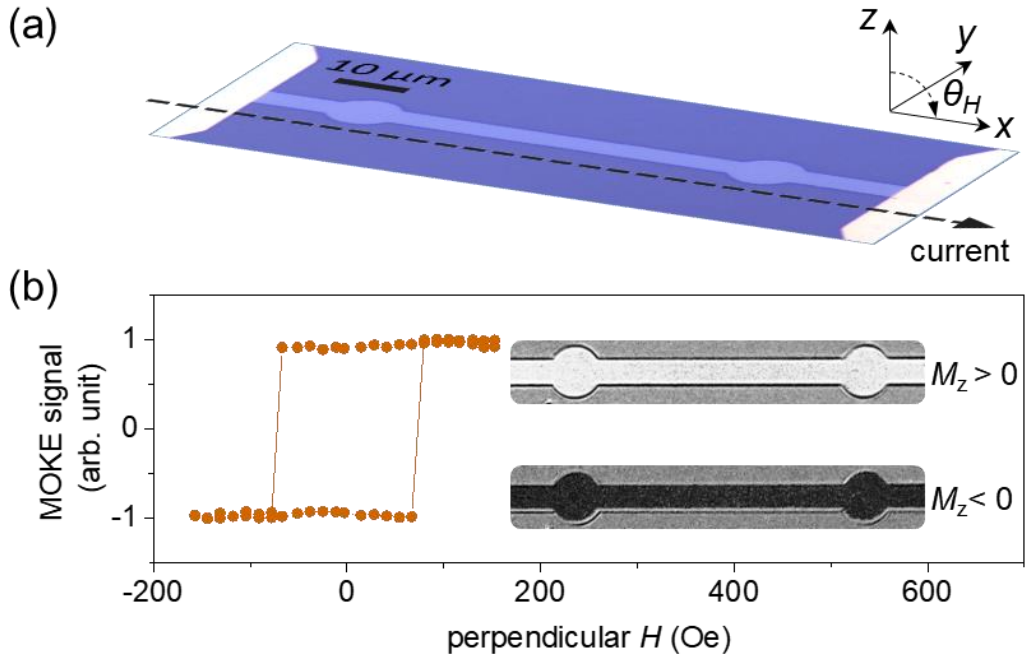

**Figure 1.** (a) The anti-notched microwire consisting of a Ta/CoFeB/MgO structure. The current is applied along the microwire and magnetic field is applied in the *xz*-plane with a polar angle of $\theta_H$. (b) Polar MOKE signal from the anti-notched microwire under a perpendicular field where the values of +1 and –1 represent averaged out-of-plane magnetization along the +z-direction ($M_z > 0$) and –z-direction ($M_z < 0$), respectively. The two insets show differentiated MOKE images with $M_z > 0$ (light contrast) and $M_z < 0$ (dark contrast).

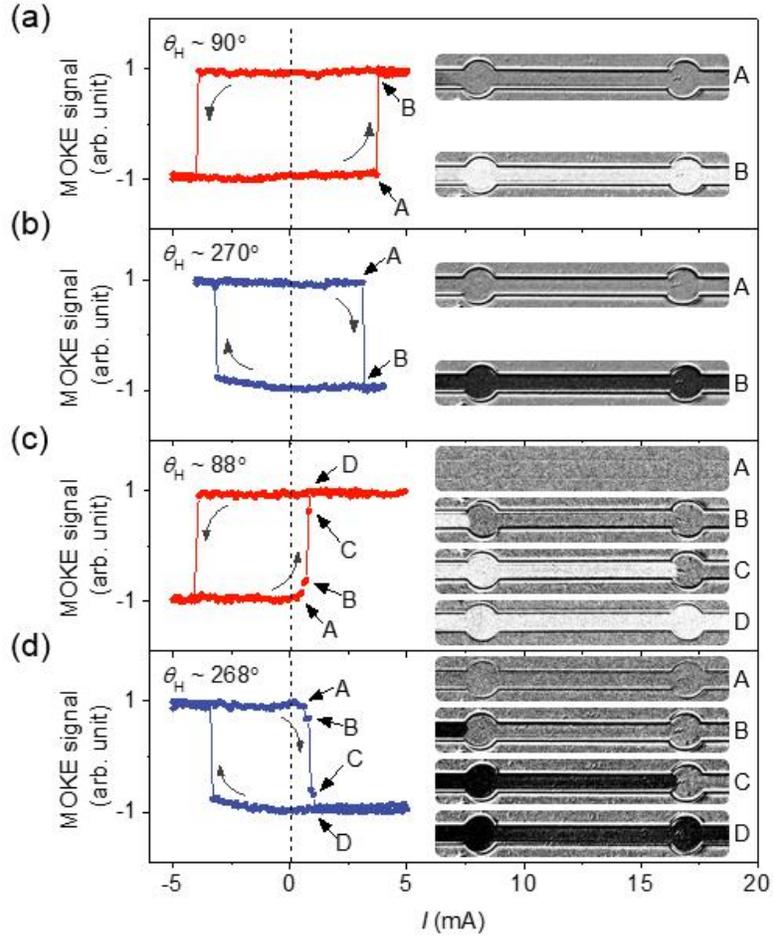

**Figure 2.** MOKE signals from the anti-notched microwire as a function of applied currents under the magnetic field with a polar angle ($\theta_H$) of ~ 90° (a), ~ 270° (b), ~ 88° (c), and ~ 268° (d). The magnetic field is fixed at 495 Oe. The A-D in the insets label the applied current values, where MOKE images are captured for illustrating the magnetic domains under the applied current values. The curved arrows represent the current sweep direction.

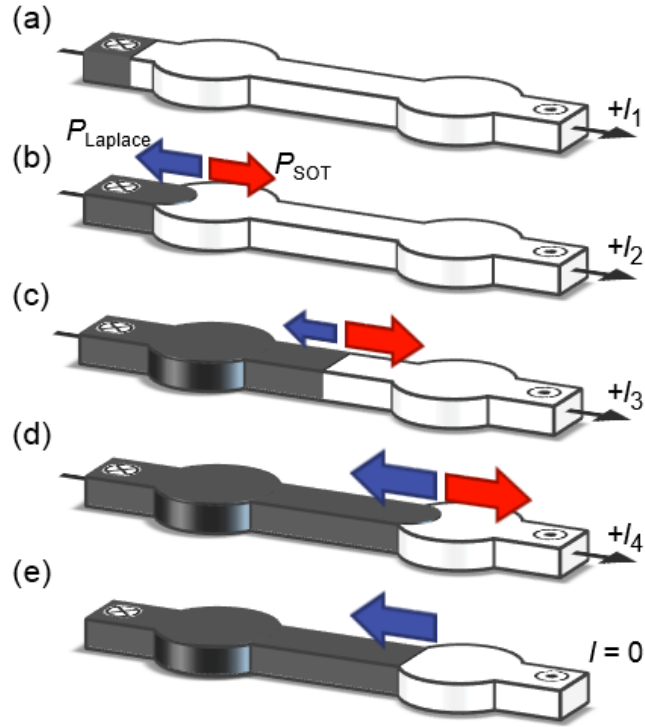

**Figure 3.** Schematic illustrations of magnetic domains under the current-induced spin-orbit torque (a-d) and after terminating the current (e). The pressure of SOT is along the current direction (red arrow). The pressure of the DW surface energy is along the opposite direction (blue arrow). An external magnetic field is applied against the current flow direction in all cases. (a) Under $I_1$, the switching process is initiated by the reverse domain nucleation at the left end of the wire. (b) DW is stabilized in a semicircular form due to a Laplace pressure at $I_2$. (c) $P_{SOT}$ overcomes $P_{Laplace}$ leading to DW expansion. (d) DW is pinned again at the next neck part due to $P_{Laplace}$. (e) Pinned DW at the neck part loses its semicircular form and becomes flat upon removing $I$.

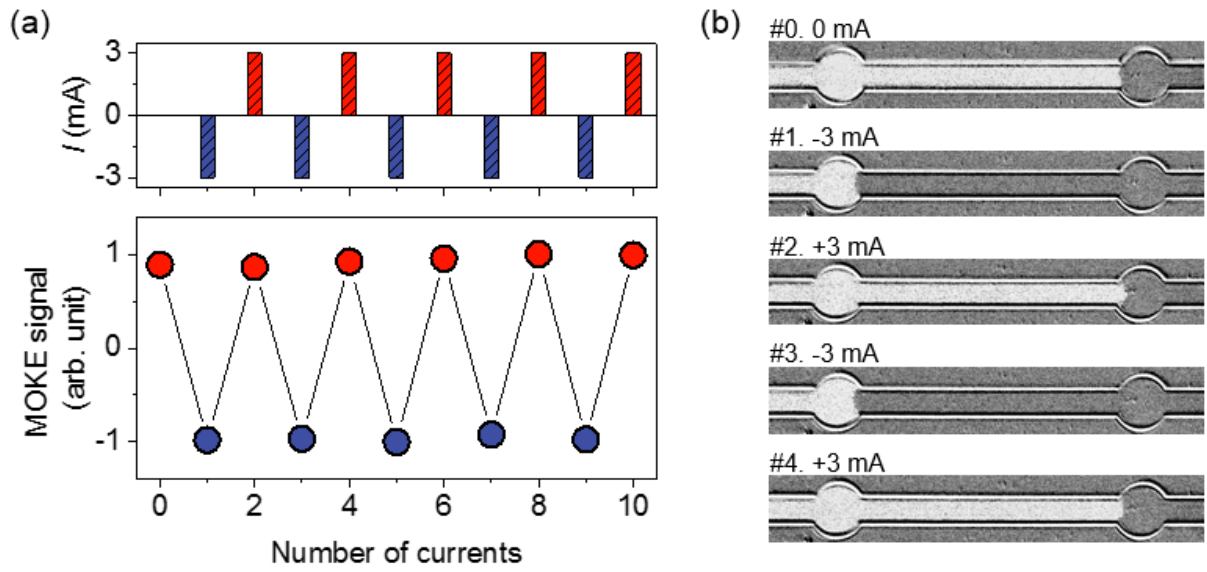

**Figure 4.** (a) Demonstration of a field-free SOT switching from repeated bipolar currents of +3 mA and –3 mA. (b) Corresponding differentiated MOKE images under number of currents. The current is not applied for the initial measurement ($I = 0$).

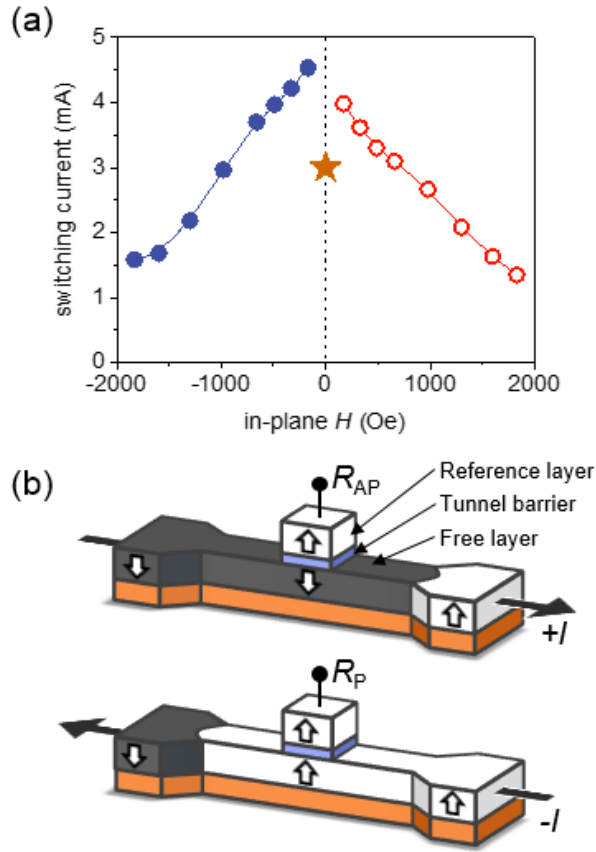

**Figure 5.** (a) Switching currents as a function of in-plane field in conventional SOT switching (filled and open circles). Star corresponds to the switching current for our field-free SOT switching (the minimum switching current is ~2 mA). (b) Schematic diagram of a field-free three terminal magnetic memory composed of anti-notches and a magnetic tunnel junction.

## ASSOCIATED CONTENT

**Supporting Information**

Additional information on the estimations of the Dzyaloshinskii-Moriya interaction constant in Ta/CoFeB/MgO, estimation of the de-pinning current from anti-notches, estimations of the de-pinning current from the DW surface energy, and nucleation of reversed domain wall. The Supporting Information is available free of charge via the Internet at http://pubs.acs.org.

## AUTHOR INFORMATION

**Corresponding Author**

*E-mail: eleyang@nus.edu.sg

**Author Contributions**

J.M.L. and H.Y. initiated the project. J.M.L., G.Y., K.C., and P.H., deposited films and fabricated devices. J.M.L., G.Y., K.C., Y.L., and R.R. performed measurements. All authors discussed the results. J.M.L., R.R., and H.Y. wrote the manuscript. H.Y. supervised the project.

**Notes**

The authors declare no competing financial interest.

†These authors contributed equally to this work.

## ACKNOWLEDGMENTS

This work was supported by the National Research Foundation (NRF), Prime Minister's Office, Singapore, under its Competitive Research Programme (CRP Award No. NRFCRP12-2013-01).

**For TOC Graphic only**

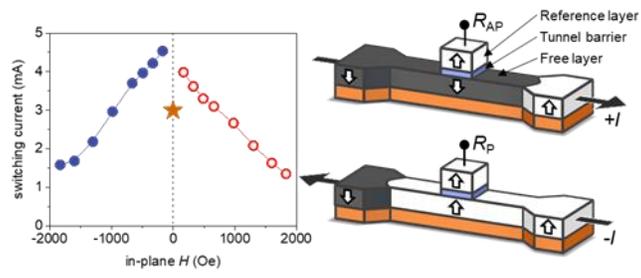